\documentclass[12pt,preprint]{aastex}

\shorttitle{Does the variation of solar inter-network horizontal field follow sunspot cycle? }
\shortauthors{Jin \& Wang}

\begin{document}

\title{Does the variation of solar inter-network horizontal field follow sunspot cycle?}

\author{C. L. Jin \& J. X. Wang}
\affil{Key Laboratory of Solar Activity, National Astronomical Observatories,
 \ Chinese Academy of Sciences, Beijing 100012, China (cljin@nao.cas.cn)}

\begin{abstract}
The ubiquitousness of solar inter-network horizontal magnetic field has been revealed by the space-borne observations with high spatial resolution and polarization sensitivity. However, no consensus has been achieved on the origin of the horizontal field among solar physicists. For a better understanding, in this study we analyze the cyclic variation of inter-network horizontal field by using the spectro-polarimeter observations provided by Solar Optical Telescope on board Hinode, covering the interval from 2008 April to 2015 February. The method of wavelength integration is adopted to achieve a high signal-to-noise ratio. It is found that from 2008 to 2015 the inter-network horizontal field does not vary when solar activity increases, and the average flux density of inter-network horizontal field is 87$\pm$1 G, In addition, the imbalance between horizontal and vertical field also keeps invariant within the scope of deviation, i.e., 8.7$\pm$0.5, from the solar minimum to maximum of solar cycle 24. This result confirms that the inter-network horizontal field is independent of sunspot cycle. The revelation favors the idea that a local dynamo is creating and maintaining the solar inter-network horizontal field.

\end{abstract}


\section{Introduction}
As the weakest magnetic field of the Sun, solar inter-network filed consists of vertical and horizontal field components. The inter-network vertical field has been found since 1975 (Livingston \& Harvey 1975; Smithson 1975), but until the late 1980s the possible presence of inter-network horizontal field was suggested by analysing the visible inter-network longitudinal field from solar disk center to limb (Martin 1988). In fact, the arcsecond scale (typically 1-2 arcseconds or smaller), short-lived (lasting lifetime of about 5 minutes) horizontal field was firstly discovered in 1996 based on the observations from Advanced Stokes Polarimeter (Lites et al. 1996). With the very sensitivity SOLIS vector spectromagnetograph (Keller et al. 2003), Harvey et al. (2007) deduced the ubiquitous ``seething" horizontal field throughout the solar inter-network region. The further evidence of inter-network horizontal field is provided by the analysis of the space-borne observations of Solar Optical Telescope (SOT; Tsuneta et al. 2008a; Suematsu et al. 2008; Shimizu et al. 2008; Ichimoto et al. 2008) on board Hinode (Kosugi et al. 2007). The Spectro-polarimeter (SP) observations have revealed that the inclination angle of inter-network magnetic field has a peak distribution at 90 degrees (Orozco Su\`{a}rez et al. 2007; Jin et al. 2012), which corresponds to the horizontal field. The horizontal field primarily concentrates in the patches on the edges of granule (Lites et al. 2008; Jin et al. 2009a), and has smaller spatial scale than the solar granule (Ishikawa et al. 2008; Jin et al. 2009a).

With the ubiquitous appearance all over the Sun including the polar regions (Tsuneta et al. 2008b; Jin \& Wang 2011), solar inter-network horizontal field contributes 10$^{25}$ Mx magnetic flux to solar photosphere per day (Jin et al. 2009b), which is one order of magnitude lower than the contribution from solar inter-network vertical field (~10$^{26}$ Mx; Zhou et al. 2013) but three orders of magnitude higher than that of ephemeral active region (~10$^{22}$ Mx; Harvey et al. 1975). Moreover, the horizontal field may also contributes to the hidden turbulent flux suggested by the studies involving Hanle depolarization of scattered radiation (Lites et al. 2008). In addition, the magnetic energy provided by the horizontal magnetic field to the quiet Sun is comparable to the total chromospheric energy loss and about ten times of the total energy loss of the corona (Ishikawa \& Tsuneta 2009). \textbf{According to the numerical simulation of quiet magnetism, Rempel (2014) found that 50\% of magnetic energy resides on scales smaller than about 0.1 arcsec, which means more magnetic energy still hidden. Based on the numerical simulations, Steiner et al. (2008) pointed that the inter-network horizontal field gets pushed to the middle and upper photosphere by overshooting convection, where it forms a layer of horizontal field of enhanced flux density, reaching up into the lower chromosphere. The inter-network horizontal field might plays a role in the atmospheric heating.}

The identification and critical importance of inter-network horizontal field have encouraged active studies to understand why so much inter-network horizontal flux is generated and what is the ultimate origin of horizontal magnetic field. The radiative magnetohydrodynamic simulations display that a local dynamo can produce the horizontal field in the quiet Sun (e.g., Sch\"{u}ssler \& V\"{o}gler 2008). The similar appearance rates of horizontal field in quiet region and plage region discovered by Ishikawa \& Tsuneta (2009) suggest that a common local dynamo which is independent of global dynamo produces the horizontal field. Furthermore, the fraction of selected pixels with polarization signals shows no overall variation by studying the long-term observations (Buehler et al. 2013). However, Stenflo (2012) argued that the global dynamo is still the main source of magnetic flux even in the quiet Sun.

\textbf{The distinction between the global dynamo which creates sunspot cycle and local dynamo which is independent of sunspot cycle would provide supporting evidence for the origin of solar inter-network horizontal field.} The long-term SP observations with high spatial resolution and polarization sensitivity provide us the opportunity to carry out this study. In order to improve the signal-to-noise ratio in weak field region, we adopt the wavelength-integration method to extract the horizontal magnetic signals observed from 2008 April to 2015 February, roughly from the solar minimum to the maximum of the current cycle.

The next section is devoted to describe the observations and data analysis. In Section 3, we present the result of cyclic invariance of inter-network horizontal field, and discuss some possible uncertainties of our result.

\section{Observations and data analysis}

The SOT/SP measurements provide high spatial resolution and polarization sensitivity observations since 2006 November. These observations were recorded by 112 wavelength points covering the spectral range from 630.08 nm to 630.32 nm, including the entire information of Stokes parameters (I, Q, U, and V) in two magnetic sensitivity Fe {\sc ii} lines. \textbf{However, because of the telemetry problems, dual mode images were unavailable after 2008 January. In order to keep consistency of image quality, the present study utilizes the observations taken after 2008 January because the uniform single mode images have been adopted since then. Avoiding those regions that were close to solar limb and polar region, a total of 430 SP measurements of quiet Sun are selected in this study, which covers the period from 2008 April to 2015 February. In order to compare the inter-network horizontal fields in different magnetic environments, 146 active regions and 109 plage regions are also selected in the same period. The exposure time of all these magnetograms is 3.2 s.}

\textbf{In order to enhance the sensitivity of weak polarization signals in the presence of measurement noise, the linear polarization signal is extracted based on the method of wavelength-integration (Lites et al. 2008). The method avoids the problems of non-convergence and non-uniqueness that arise in the inversions of noisy profiles, and provides the optimal sensitivity to the weak polarization signal.}

\textbf{Just as accounted by Buehler et al. (2013) and Lites et al. (2014), the rms contrast of the Stokes I is not constant but shows a fluctuation due to the temperature fluctuation on the spacecraft during the long-term observations. In order to examine the instrumental effect, we first show the rms intensity contrast as a function of the means distance from solar disk center, which is shown in the left panel of Figure 1. An obvious dropping of the rms intensity contrast is found when the observed regions were far away from solar disk center. However, within the distance of 0.2 solar radius, the changing of intensity contrast is not obvious. To examine the long-term variation of rms contrast, we selected all quiet magnetograms from SP measurement within the distance of 0.2 solar radius, and analyzed the rms contrast in the period of investigation. The result is shown in the middle panel of Figure 1. The fluctuation reaches as much as 1.1\%, confirming the result of Buehler et al. (2013) and Lites et al. (2014). To reduce the fluctuation, we degrade all observations spatially by a 3$\times$3 smoothing average, and the result is shown in the right panel of Figure 1. It can be found that the average fluctuation has a slight decline, falling down to 0.7\% after smoothing, while the average rms contrast drops to 5.23 from original value of 6.99. Furthermore, during the investigation, the rms contrast does not show regular variation but a fluctuation around a constantly horizontal line with time. Therefore, we do not make further instrumental amendment except the spatial smoothing.}

\textbf{In this study, not all observations are located on the disk center, so the center-limb effect of magnetic field needs to be considered. The limb-weakening of circular polarization has been identified (e.g., Jin \& Wang 2011; Lites et al. 2014). However, there is still few analysis on the center-limb effect of the linear polarization. Here, we adopt two methods to analyze the variation of linear polarization when the observed regions were far away from solar disk center. On the one hand, we assume that there are only two possible cyclic variations of the horizontal field in quiet Sun: linear correlation (or anti-correlation) with sunspot cycle and keeping constant in the sunspot cycle. Based on the assumption we analyze these magnetograms of quiet Sun observed from 2008 April to 2009 July, an interval of invariant sunspot number. The result is shown in the top panel of Figure 2. We can find that moving far away from solar disk center, the horizontal field in quiet Sun does not display obvious changes. On the other hand, we avoid the observation from solar polar region, and study these local observations of quiet Sun from solar disk center to limb in a few days, i.e., a period from 2008 December 29 to 2009 January 4. These magnetograms locate at almost the same magnetic environment of the Sun's disk, resembling a series of observations from solar center to limb at the same time. The corresponding result is shown in the bottom panel of Figure 2. For the magnetic distribution from solar disk center to limb, the variation of horizontal field is still not obvious. Therefore, in this study, the magnitude of horizontal field is considered as independent of observational location on the solar disk.}

The pixel area in each magnetogram is corrected by $\cos(\alpha)$, where $\alpha$ is the heliocentric angle. We identify the horizontal magnetic structures by setting the thresholds of 0.2 Mm$^{2}$ on the area (i.e., equivalent to the area of 4 pixels) and 1.5 times of the noise level, where the magnetic noise of horizontal field is about 50 G. Note that, unlike the vertical field $B_{v}$, the horizontal field $B_{h}$ times the pixel area is not the measurement of magnetic flux since the field is transverse to the line-of-sight. We first adopt the equivalent spatial scale of the horizontal magnetic structures in the transverse surface, i.e., $d=\sqrt{S/\pi}\times2$, where $S$ is the area of horizontal magnetic structure in the transverse surface. Then we assume a vertical height of $h=100$ km (Lites et al. 1996; Jin et al. 2009b), which is comparable to the photospheric scale height, and compute the area of horizontal magnetic structures by $d\times h$. Finally we obtain the magnetic flux of these horizontal magnetic structures by considering their average flux density and area. In order to avoid the effect from the solar network field and plage regions as well as active regions, we exclude those horizontal magnetic structures with larger spatial scale and stronger magnetic flux density by considering the magnetic flux larger than 5.0$\times10^{17}$ Mx. The identification of IN region is displayed in Figure 3. The region within the red contours represents the horizontal magnetic structures in the active region, plage region as well as network region. The average area ratio of inter-network region to the entire SP quiet magnetogram reaches 98.0\%, and exceeds 42.7\% even through in the measurement including active region.

\section{Cyclic variation of inter-network horizontal field}

In this study, we set the threshold of 1.5 times of the noise level on the inter-network horizontal magnetic field to analyze its flux density. On average 17.7\% of pixels in a quiet map carry the significant horizontal magnetic signal, and the average area ratio of significant horizontal magnetic signals also reaches 13.7\% in a magnetogram including active region. These magnetograms are also analyzed by using different thresholds, such as the 2 and 2.5 times of noise level, but the results display no obvious dependence of the threshold choice. Higher thresholds severely reduce the area occupancy of magnetic signals and enhance the magnetic flux density, and result in significantly poor statistics. Because our results are not affected, description of employing higher thresholds is not discussed in this study.

We first compute the horizontal magnetic flux density of each quiet magnetogram for the identified solar inter-network region, and obtain the monthly average distribution and the standard deviation of inter-network horizontal field, which is shown by the black symbol in the top panel of Figure 4. The corresponding monthly average sunspot number in this period is displayed in the bottom panel of Figure 4. The average inter-network horizontal flux density reaches 87$\pm$1 G. Comparing the inter-network vertical flux density above the 1.5 times of noise level, the obvious imbalance between vertical and horizontal flux densities is confirmed. Furthermore, it can be found that the imbalance is invariable, i.e., a constant of 8.7$\pm$0.5, during the ascending phase of solar cycle 24, which is shown in the middle panel of Figure 4. Within the scope of deviation, the distribution of inter-network horizontal field does not vary from the solar minimum to the current maximum of cycle 24 either. The fact of no cyclic variation of inter-network horizontal field suggests that it is predominantly governed by a process that is independent of the global solar cycle.

\textbf{Due to the randomness of SP local observations, it is necessary to examine whether the local information of horizontal field can represent its full disk information. We choose two target regions: inter-network regions close to an active region and within a large-scale quiet region. The two target regions perhaps suffer from different influences from either strong or weak surrounding flux distribution by some types of flux diffusion or magnetic interaction. Based on the consideration, we also analyse the inter-network horizontal measurements surrounding active region or plage region. The result is shown by the green symbol in the top panel of Figure 4. It can be found the IN horizontal field surrounding the active region does not display the cyclic variation, either. Comparing the cyclic variations of inter-network horizontal field in the large-scale quiet region and at the surroundings of active or plage regions, we can find that their variational ranges are almost the same. Just as the cyclic behavior of longitudinal field (Jin et al. 2011), the diffusion of active region is likely to make an obvious effect only on the strong network field. The diffusion of active region seems to be little and even insignificant for the inter-network field.}

Our results also indirectly confirm early findings obtained by Ito et al. (2010) and Shiota et al. (2012) who found the invariance of the inter-network magnetic field in the solar polar region by analysing Hinode/SOT data. Some detailed analysis of the solar inter-network region has strongly suggested the presence of local dynamo (Lites 2011; Buehler et al. 2013; Lites et al. 2014). Furthermore, the numerical simulation of local dynamo can reproduce the process of generating the solar inter-network horizontal field (V\"{o}gler \& Sch\"{u}ssler 2007; Sch\"{u}ssler \& V\"{o}gler 2008). Such a local dynamo process seems to naturally explain the cyclic invariance of inter-network horizontal field.

The conclusion still needs to be examined by the observations with large field-of-view and continuous temporal coverage. In a sense the partition of network and inter-network regions only by the magnitude of magnetic flux is of uncertainties. The magnetic flux of network magnetic structures in the rapid evolving phase is sometimes very weak, and the magnetic flux of some inter-network magnetic structures is even larger than that of network magnetic structures (Wang et al. 1995; Zhou et al. 2013). The observations with full-disk coverage and high spatial resolution and polarization sensitivity are still needed to examine our result.

\acknowledgments

The authors are grateful to the team members who have made great contribution to the Hinode mission. Hinode is a Japanese mission developed and launched by ISAS/JAXA, with NAOJ as domestic partner and NASA and STFC (UK) as international partners. It is operated by these agencies in co-operation with ESA and NSC (Norway). The work is supported by the National Basic Research Program of China (G2011CB811403) and the National Natural Science Foundation of China (11003024, 11373004, 11322329, 11221063, KJCX2-EW-T07, and 11025315).

\begin{figure}
\includegraphics[scale=1]{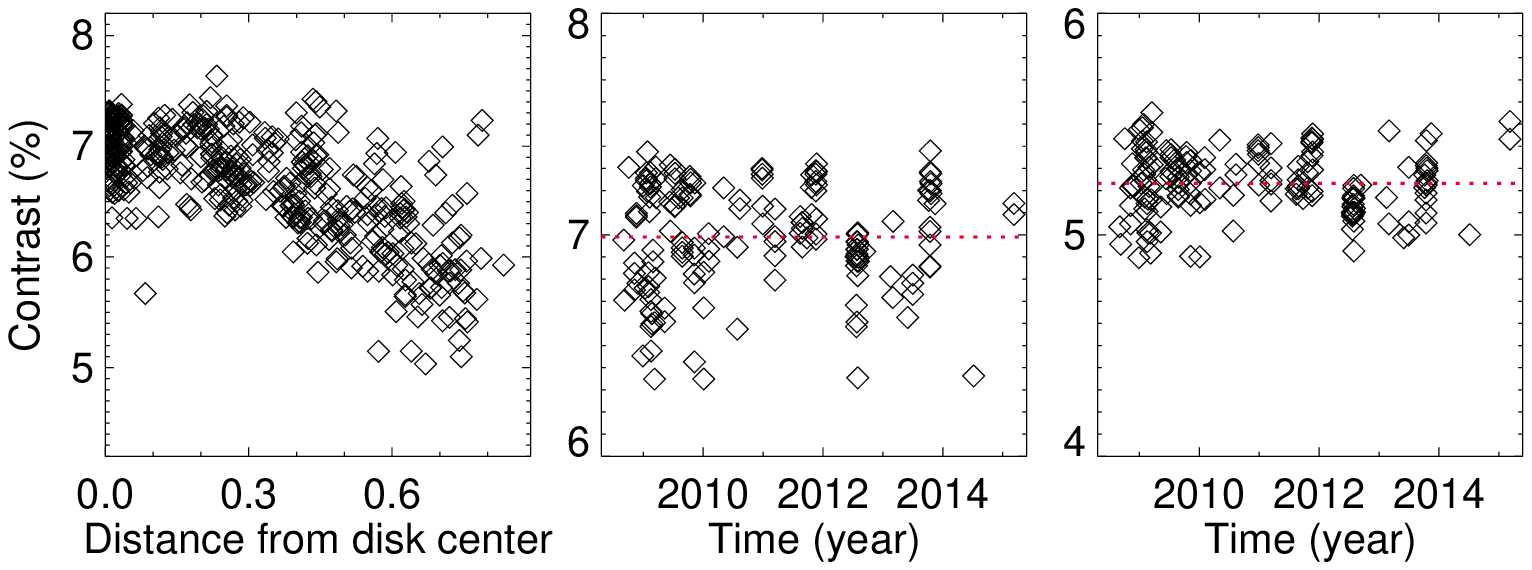}
\caption{\textbf{Properties of the rms intensity contrast for SOT quiet magnetograms. Left panel: the rms contrast as a function of distance from solar disk center. Middle panel: the original variation of rms contrast for these quiet magnetograms within 0.2 solar radius. Right panel: the variation of rms contrast for these continuum images after the spatial smooth.}}
\end{figure}

\begin{figure}
\includegraphics[scale=0.8]{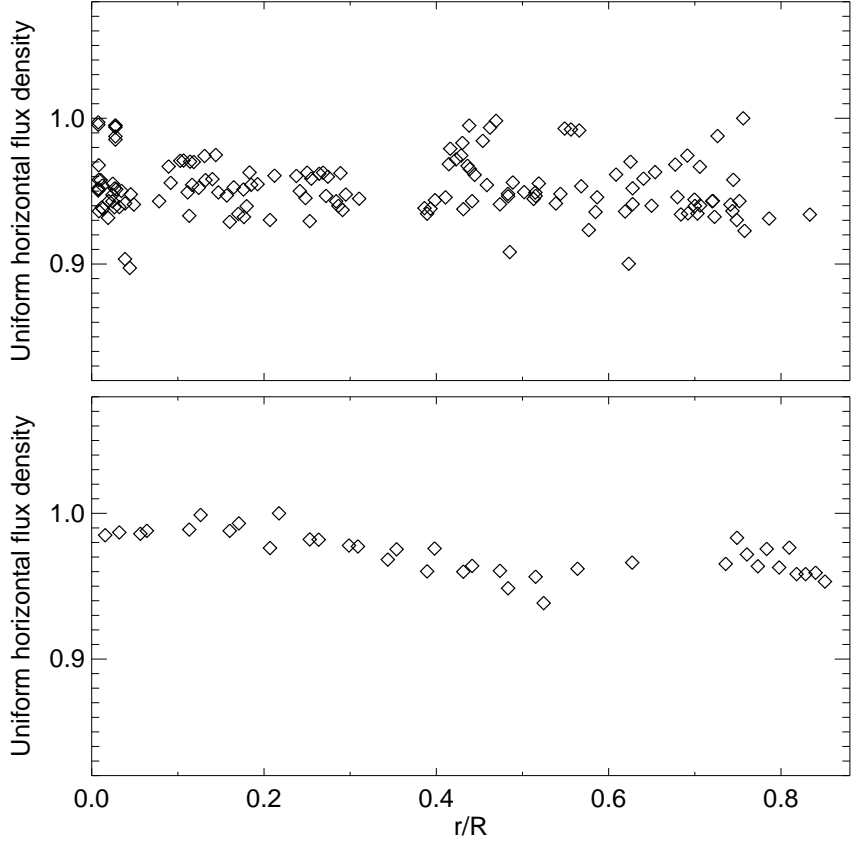}
\caption{\textbf{The center-limb variation of horizontal field in quiet regions based on two different methods. By assuming two possible cyclic variations of horizontal field in quiet region (e.g., the linear correlation or anti-correlation with sunspot cycle and keeping constant in the sunspot cycle), the quiet magnetograms observed from 2008 April to 2009 July are used to analyze the center-limb variation due to invariant sunspot number in this period, which is shown in the top panel. By observing the different local quiet regions almost at the same time and in the same full-disk magnetic environment, the center-limb variation of horizontal field is analyzed, which is shown in the bottom panel.}}
\end{figure}

\begin{figure}
\includegraphics[scale=0.8]{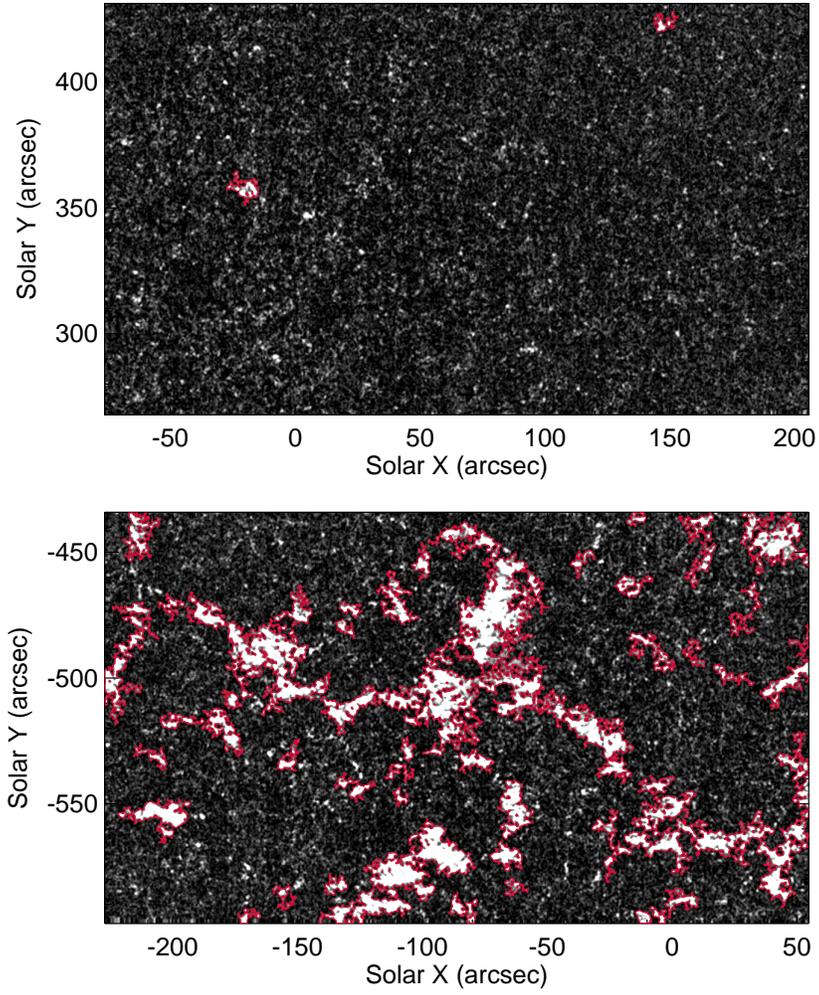}
\caption{Two samples of identifying inter-network region in the horizontal magnetograms.}
\end{figure}

\begin{figure}
\includegraphics[scale=0.8]{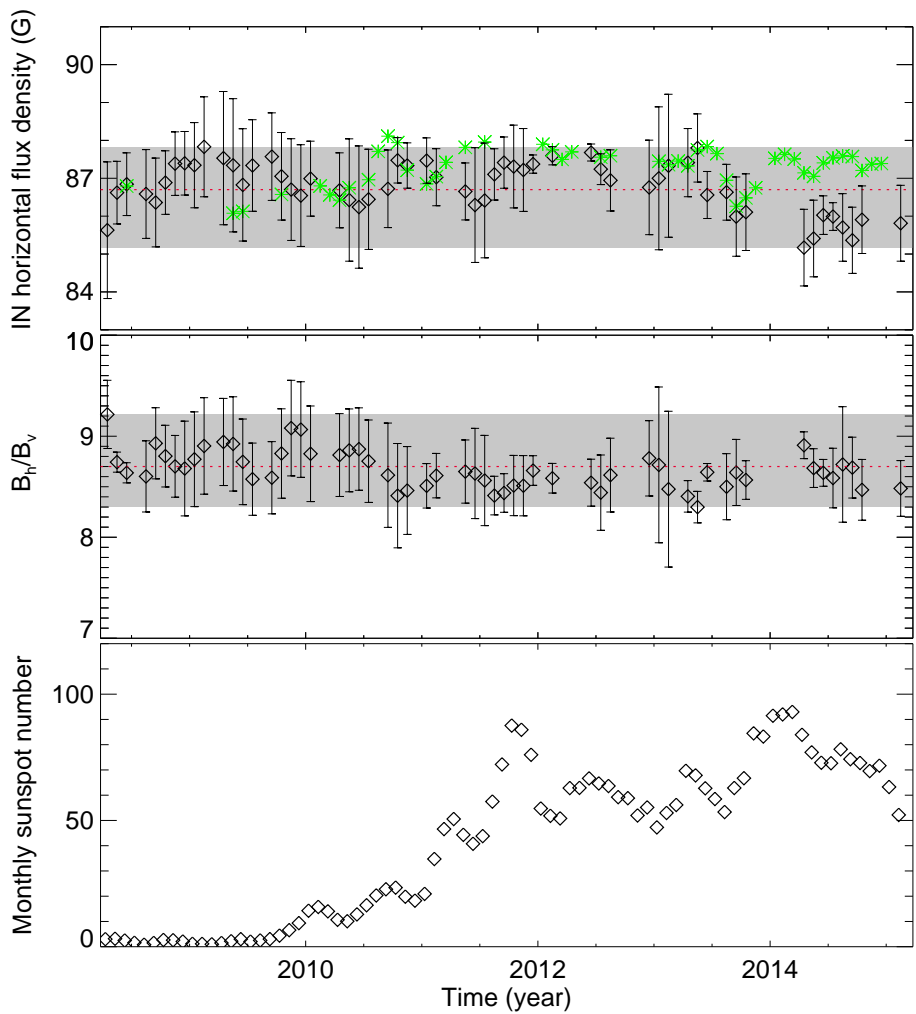}
\caption{The cyclic variations of solar inter-network horizontal magnetic field (top panel) and the imbalance between inter-network horizontal and vertical magnetic flux densities (middle panel). The corresponding variation of monthly sunspot number is displayed in the bottom panel. In the top panel, the green symbol represents the monthly distribution of inter-network horizontal field surrounding active regions or plage regions, and the black symbol represents the inter-network horizontal field in large-scale quiet region. In the top and middle panels, the red dotted lines represent their corresponding average values in large-scale quiet background, and the gray scales mean their variational ranges.}
\end{figure}

\end{document}